\documentclass[aps,pra,amsmath,amssymb,reprint]{revtex4-1}
\usepackage{dcolumn}
\usepackage{bm}
\usepackage{graphicx}
\usepackage{longtable}
\usepackage{footnote}
\usepackage{float}

\usepackage{color}

\draft 

\begin{document}

\newcommand{\bra}[1]{\langle #1|}
\newcommand{\ket}[1]{|#1\rangle}
\newcommand{\braket}[2]{\langle #1|#2\rangle}
\newcommand{\p}{^\prime}
\newcommand{\pp}{^{\prime\prime}}

\title{Universal behaviour of diatomic halo states and the mass sensitivity of their properties}


\author{A. Owens$^{1,2}$}
\author{V. \v{S}pirko$^{3,4}$}\email{vladimir.spirko@marge.uochb.cas.cz}
\affiliation{$^1$Center for Free-Electron Laser Science (CFEL), Deutsches Elektronen-Synchrotron DESY, Notkestrasse 85, 22607 Hamburg, Germany}
\affiliation{$^2$The Hamburg Center for Ultrafast Imaging, Universit\"{a}t Hamburg, Luruper Chaussee 149, 22761 Hamburg, Germany}
\affiliation{$^3$Institute of Organic Chemistry and Biochemistry, Flemingovo n\'{a}m.~2, 166 10 Prague 6, Czech Republic}
\affiliation{$^4$Department of Chemical Physics and Optics, Faculty of Mathematics and Physics, Charles University in Prague, Ke Karlovu 3, CZ-12116 Prague 2, Czech Republic}

\date{\today}

\begin{abstract}
The scattering and spectroscopic properties of molecular halo states can serve as sensitive probes of the constancy of the electron-to-proton mass ratio $\beta=m_e/m_p$. Since halo states are formed by resonant $s$-wave interactions, their properties exhibit universal correlations that are fairly independent of the interactions at short distances. For diatomic molecules, these properties depend on a single parameter only, and so this `universality' means that all the characteristics of a diatomic halo state can be determined with high precision if only one parameter is accurately known. Furthermore, this knowledge can be used to establish the respective property mass sensitivities for investigating the stability of $\beta$. Here, we show for the halo states of the helium dimers that the relationship between the probed properties and their mass sensitivity can be derived from numerically exact solutions of suitable radial Schr\"{o}dinger equations for a set of effective potential energy curves. The resulting relations exhibit a weak dependence on the short-range part of the used potentials and a near-negligible dependence on the `higher-order' nonadiabatic, relativistic, QED and residual retardation effects. The presented approach is thus a robust alternative to other literature approaches, particularly in cases where a lack of experimental data prevents an accurate interaction potential from being determined.
\end{abstract}

\maketitle

\newpage

\section{Introduction}

 The formation of a quantum halo state is a phenomenon associated with loosely bound particles held in short-range potential wells. Halo states extend over an unusually large space and many of their properties hinge on the tail of their wavefunctions exhibiting unusual scattering properties~\cite{JensenAS}. For instance, they allow for the formation of three body systems exhibiting the Efimov effect~\cite{Naidon}; the long range asymptote of the wavefunction of such a state implies quite distant correlations which are relevant for the formation of Bose-Einstein condensates~\cite{Yukalov} and sympathetic cooling~\cite{Barletta}. The actual scattering is characterised by an $s$-wave scattering length that is much larger than the range of the particle interactions. It is therefore conceivable that the scattering length $a$ and its related properties, such as the binding energy $D_0$ or the average value of the internuclear separation $\langle R \rangle$, can exhibit highly anomalous mass dependencies, thus serving as promising probes of a possible variation of the electron-to-proton mass ratio $\beta=m_e/m_p$. Interestingly, as noted by Abraham et al.~\cite{Abraham} and analyzed in detail by Chin and Flambaum~\cite{Chin}, this dependence can be dramatically enhanced in collisions of atoms near narrow Feshbach resonances. 

 The development of laser-cooling techniques enables long-range halo states to be studied with very high spectral resolution, for example, using photoassociation spectroscopy~\cite{Jones}. Moreover, as was recently shown for He$_2$ in its ground electronic state~\cite{Zeller}, the square $|\Psi|^2$ of the wavefunction of the halo state can be measured by recording a large number of Coulomb explosion events using cold target recoil ion momentum spectroscopy (COLTRIMS)~\cite{Ullrich}. Since the properties of halo states are expected to be extremely sensitive to tiny variations of the interaction potential, we have performed calculations for the halo states of the ground and excited $^5\Sigma_g^+$ electronic states of the helium dimers, for which precise theoretical interaction potentials~\cite{PCJS,PJ} and appropriate experimental bond lengths and binding energies are available. For the electronic ground state these data were obtained by molecular beam diffraction from a transmission grating~\cite{Grisenti} and from the above mentioned Coulomb explosion experiment~\cite{Zeller}; while for the $^5\Sigma_g^+$ electronic state these data were obtained from a two-photon photoassociation experiment~\cite{Moal}.

\section{Methods}

 In our approach, the potentials are used to generate relationships between 
the scattering properties of the probed states and their actual values, from which the mass sensitivities can be determined. This is done through numerical solution of the radial Schr\"{o}dinger equation for a set of effective potentials $V^{sc}(R)$, obtained 
by scaling the interaction potential $V(R)$ with a constant scaling factor $f$ (for further details see Ref.~\cite{SSS}),
\begin{equation}
V^{sc}(R)\!=\!f\cdot V(R).
\end{equation}
The scaled potentials are then utilized to establish the $f$-dependence of the scattering properties, and by combining these dependencies, a direct relationship between the probed properties can be found. For example, establishing the two relations $D_0=D_0(f)$ and $a=a(f)$ for a set of values of $f$, e.g. $f_i$ ($i=1,2,\ldots$), allows us to obtain the pairs $\{D_0(f_i),a(f_i)\}$, which subsequently defines the one-to-one relations $a=F(D_0)$.

 Calculations were performed using the effective vibrational Hamiltonian of Herman and Asgharian~\cite{HA} for nuclear motion in $^1\Sigma$ state molecules,
\begin{eqnarray} H_{\rm eff}\!=\!-\frac{\hbar^2}{2m}\frac{\rm d}{{\rm d}
R}\Biggl(\!1\!+\!\beta g_v(R)\!\Biggr)\frac{\rm d}{{\rm d}R}\!+\!V_{\rm ad}(R)\!+\!V^{'}(R),
\label{eq.Ham}
\end{eqnarray}
where $m$ is the appropriate nuclear reduced mass, $V_{\rm ad}$ is the `adiabatic' part of the molecular potential energy function (assumed to include Born-Oppenheimer, adiabatic, relativistic, QED and residual retardation terms), and the terms $V^{'}(R)$ and $g_v(R)$ account
for nonadiabatic effects. The so-called vibrational $g_v$-factor is fixed to its {\it ab initio} value, and the effective potential energy function $V=V_{ad}(R)+V^{'}(R)$ is determined either from first principles or from fitting to the available experimental data.

 The mass sensitivity of the probed properties, for example, the binding energy $D_0$ and the scattering length $a$, are described by the following expressions,
\begin{equation}
K_{\beta}=\frac{\beta}{D_0}\frac{{\rm d}D_0}{{\rm d}\beta},
\label{eq:K}
\end{equation}
\begin{equation}
T_{\beta}=\frac{\beta}{a}\frac{{\rm d}a}{{\rm d}\beta}.
\label{eq:T}
\end{equation}
The resulting sensitivity coefficients, $K_{\beta}$ and $T_{\beta}$, can then be used to determine the induced shift of the respective property,
\begin{equation}
\frac{\Delta\nu}{\nu_0}=P_{\beta}\frac{\Delta\beta}{\beta_0} \qquad (P=K,T, ..),
\label{eq.shift}
\end{equation}
where $\Delta\nu=\nu_{\mathrm{earlier}}-\nu_0$ is the change in the property, and
$\Delta\beta=\beta_{\mathrm{earlier}}-\beta_0$ is the change in $\beta$, both with respect to their
present day values $\nu_0$ and $\beta_0$.

 Alternative approaches for relating the scattering properties of diatomic molecules do exist, however, these methods have certain drawbacks because of the assumptions they employ: The analytical formulas based on semiclassical and quantum defect theories, and specific $-C_n/R^n$ asymptotes of interaction potentials (see, e.g., Refs.~\cite{Gribakin,Marinescu,Szmytkowski,Boisseau,Gao,Gao:2004}), are quantitative only in the case of very large scattering lengths (see below). Whereas the spectroscopic approach, which constructs potentials from spectroscopic data and then extracts the sought information from the relevant wavefunctions, 
or from extrapolating the phases of the last bound levels towards the dissociation limit (see, e.g., Refs.~\cite{Crubellier,SKT2008,Borkowski}), may be hampered by the dependence of the least bound state on the interaction potential; for example, a tiny variation of $V(R)$ may introduce or remove a bound state causing the value of $a$ to pass between $\pm\infty$ as a result (see, e.g., the studies on $^6$Li~\cite{Crubellier} and $^{84}$Sr$^{88}$Sr~\cite{SKT2008}). 


\section{Results and Discussion}

\begin{figure*}[ht]
\label{fig:fig1}
\includegraphics{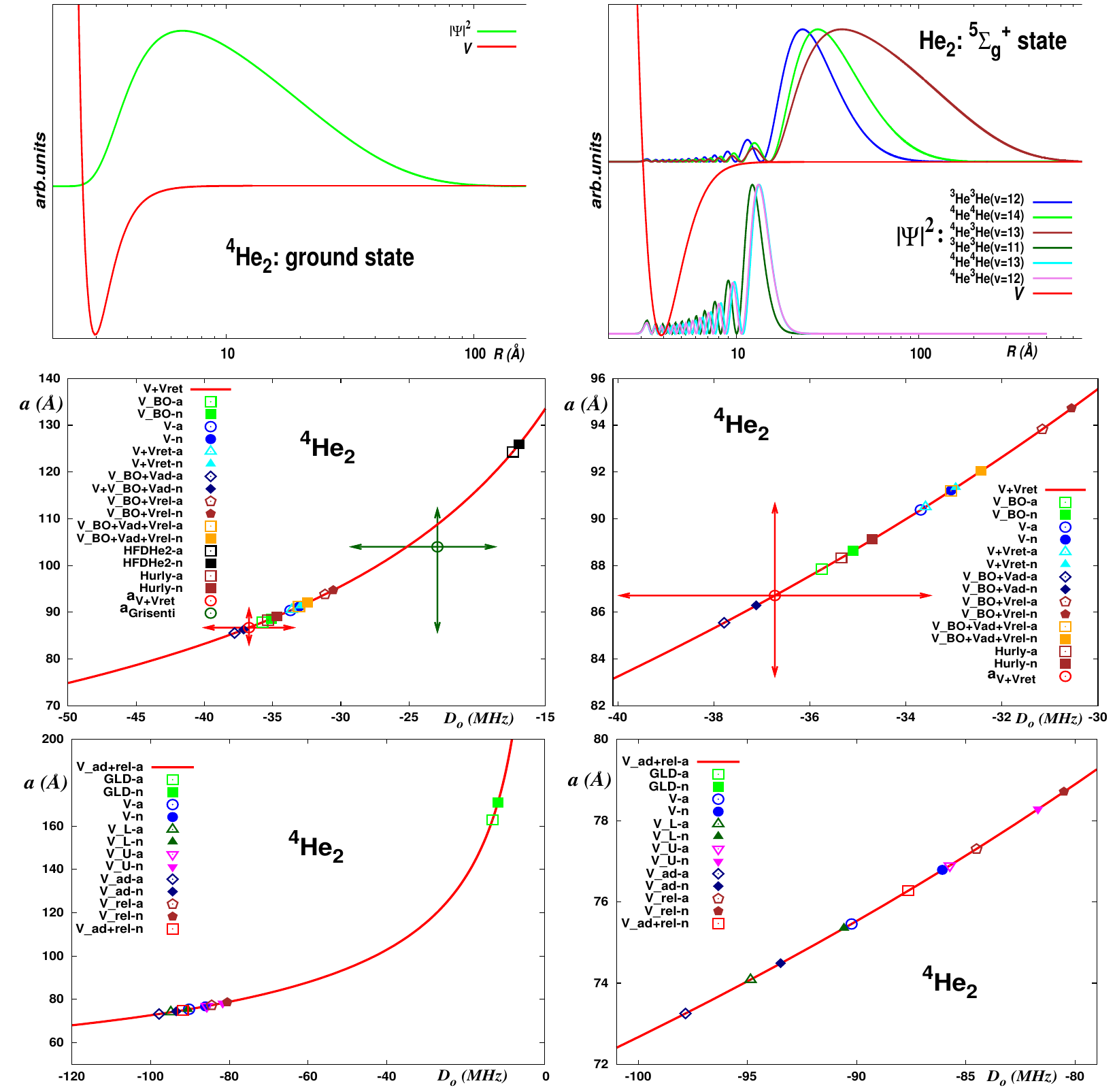}
 \caption{Top panels: The potential energy function ($V$) and square of the wavefunction ($|\Psi|^2$) of the highest vibrational states of the helium dimers. The interaction potentials are from Refs.~\cite{PJ,PCJS}. Middle panels: The scattering length ($a$) vs.\ binding energy ($D_0$)
for the bound state of $^4$He$_2$ in the ground electronic state. The curve was obtained by scaling the $V\!+\!V_{\rm ret}$ theoretical potential of Ref.~\cite{PCJS}. $a_{V\!+\!V_{\rm ret}}$ and $a_{\rm Grisenti}$ represent the values and error bars derived using the latter theoretical potential and
from experiment~\cite{Grisenti}, respectively. The points were obtained using the remaining theoretical potentials of Ref.~\cite{PCJS} and the empirical `HFDHe2' and `Hurly' potentials
of Refs.~\cite{Aziz,Hurly}. The calculations were performed using the atomic (suffix a) and nuclear (suffix n) masses. Bottom panels: $a$ vs.\ $D_0$ for the highest bound state of $^4$He$_2$ in the $^5\Sigma_g^+$ electronic state. The curves were obtained by scaling the $V\!+\!\delta\!V_{\rm ad}\!+\!\delta\!V_{\rm rel}$ theoretical potential of Ref.~\cite{PJ}. The points were obtained using the remaining theoretical potentials of Ref.~\cite{PJ} and Ref.~\cite{GLD} (denoted GLD).
}
\end{figure*}

 The ground state potential of Ref.~\cite{PCJS}, denoted PCJS, supports a bound state only for the heaviest stable isotopomer $^4$He$_2$. As seen in the top left panel of Figure 1, this state is a true halo state. The much deeper potential of the $^5\Sigma_g^+$ electronic state from Ref.~\cite{PJ}, denoted PJ, supports a number of bound ro-vibrational states for each of the He$_2$ isotopomers. However, only the least bound vibrational states (namely $v=12$, 13 and 14 for $^3$He$_2$, $^4$He$^3$He and $^4$He$_2$, respectively) exhibit the behaviour of a quantum halo state (see the top right panel of Figure 1, and note that quantum halo states are defined as bound states of particles with a radius extending into classically forbidden regions~\cite{JensenAS}). 

 The most important helium dimer appears to be that of the ground electronic state. This stems from the role of helium in the planned redefinition of the Kelvin unit of thermodynamic temperature in terms of the Boltzmann constant~\cite{LinH}. Since the thermophysical 
properties of helium computed from first principles are more accurate than their experimental counterparts, the present thermometry relies on the theoretical values (see, e.g., Ref.~\cite{CPKMJS}). 
These thermophysical properties are defined in terms of the same interaction potential as the binding energy and scattering length, quantities which can be very accurately determined in experiment. The thermophysical properties can therefore be checked, and perhaps predicted more accurately, using the experimental values of $D_0$ and $a$ and a key aspect in this process is the interaction potential.

 The best available potential for the helium dimer was recently computed~\cite{PCJS} and possesses submillikelvin uncertainties. The potential accounts for all relevant `higher-order' effects (adiabatic, relativistic, quantum electrodynamical (QED) and retardation corrections), allowing for a direct and detailed analysis of each contribution. Calculations with these potentials, represented 
by Table 2 of Ref.~\cite{PCJS} and Table 5 of Ref.~\cite{PJ}, reveal the following important facts (note that we adopt the same notation here as the aforementioned tables): (a) the most rigorous theoretical potential predicts the binding energy within the error bars of the experimental value deduced from Coulomb explosion measurements, (b) the adiabatic and relativistic effects contribute significantly meaning that the effect of retardation is small, (c) accurately accounting for nonadiabatic effects is possible through adiabatic calculations with atomic mass values (instead of nuclear masses). 

\begin{table*}
\caption{\label{tab:scl1}Scattering lengths $a$ of $^4$He$_2$ in its ground ($^1\Sigma_g^+$) and excited ($^5\Sigma_g^+$) electronic states (in \AA).}
\vspace*{-3mm}
\begin{center}
\begin{tabular}{c @{\extracolsep{0.30in}}
                c @{\extracolsep{0.22in}}
                c @{\extracolsep{0.22in}}
                c @{\extracolsep{0.22in}}
                c @{\extracolsep{0.22in}}
                }
        \hline\hline\\[-3mm]
Potential                        & Calc-1 & Calc-2 & Calc-3& Calc-4 \\
        \hline\\[-2mm]
 &  &  \qquad\qquad\qquad $^1\Sigma_g^+$  & & \\[2mm]
$V_{\rm BO}$                         & 87.847 & 88.634 & 86.718(+3.889,-3.409)& 86.730(+3.889,-3.410) \\
$V_{\rm BO}+V_{\rm ad}$                  & 85.550 & 86.295 & 86.718(+3.889,-3.409)& 86.730(+3.889,-3.410) \\
$V_{\rm BO}+V_{\rm rel}$                 & 93.827 & 94.730 & 86.715(+3.889,-3.409)& 86.726(+3.889,-3.410) \\
$V_{\rm BO}+V_{\rm ad}+V_{\rm rel}$          & 91.194 & 92.045 & 86.715(+3.889,-3.409)& 86.726(+3.889,-3.410) \\
$V$                              & 90.376 & 91.211 & 86.716(+3.889,-3.409)& 86.728(+3.889,-3.409) \\
$V+V_{\rm ret}$                      & 90.502 & 91.339 & 86.716(+3.888,-3.409)& 86.728(+3.889,-3.409) \\[2mm]
$Hurly$                         & 88.323 & 89.119 & 86.715(+3.889,-3.409)& 86.727(+3.889,-3.410) \\
$HFDHe2$                         &124.304 &125.909 & 86.723(+3.889,-3.409)& 86.735(+3.889,-3.410) \\[2mm]
&  &  \qquad\qquad\qquad $^5\Sigma_g^+$  & & \\[2mm]
$V_U$                                      &76.893  &78.287  &75.116(18) &75.124(18) \\
$V$                                        &75.456  &76.787  &75.115(18) &75.122(18) \\
$V_L$                                      &74.082  &75.355  &75.113(18) &75.121(18) \\
$V\!+\!\delta V_{\rm ad}$                     &73.252  &74.491  &75.116(19) &75.124(18) \\
$V\!+\!\delta V_{\rm rel}$                    &77.302  &78.713  &75.115(18) &75.122(18) \\
$V\!+\!\delta V_{\rm ad}\!+\!\delta V_{\rm rel}$ &74.968  &76.278  &75.117(18) &75.124(18) \\[2mm]
$GLD$                                      &162.86  &170.92  &75.268(19) &75.276(19) \\[2mm]
    \hline\hline
\end{tabular}
\end{center}
{\footnotesize If not stated otherwise the interaction potentials and notation are from Refs.~\cite{PJ,PCJS}. $D_0=36.73$ MHz~\cite{Zeller} and 91.35 MHz~\cite{Moal}
for the ground and excited state of $^4$He$_2$, respectively.
Calc-1 and Calc-2: derived from  the `zero collision energy' wavefunctions using atomic ($m_{\rm atom}=4.00260325413$~amu) and nuclear ($m_{\rm nuc}=4.001506179125$~amu) masses, respectively; Calc-3 and Calc-4: evaluated using the `$a$ vs.\ $D_0$' relations with atomic and nuclear masses, respectively. $Hurly$: obtained using the potential of Ref.~\cite{Hurly}. $HFDHe2$: obtained using the potential of Ref.~\cite{Aziz}. $GLD$: obtained using the potential of Ref.~\cite{GLD}.}
\end{table*}

\begin{table*}
\caption{\label{tab:scl2}Average internuclear separations $\langle R\rangle$ of $^4$He$_2$ in its ground ($^1\Sigma_g^+$) and excited ($^5\Sigma_g^+$) electronic states (in \AA).}
\vspace*{-3mm}
\begin{center}
\begin{tabular}{c @{\extracolsep{0.30in}}
                c @{\extracolsep{0.22in}}
                c @{\extracolsep{0.22in}}
                c @{\extracolsep{0.22in}}
                c @{\extracolsep{0.22in}}
                }
        \hline\hline\\[-3mm]
Potential                        & Calc-1 & Calc-2 & Calc-3& Calc-4 \\
        \hline\\[-3mm]
 &  &  \qquad\qquad\qquad $^1\Sigma_g^+$  & & \\[2mm]
$V_{\rm BO}$                         & 45.802 & 46.195 & 45.237(+1.946,-1.706)& 45.243(+1.946,-1.706) \\
$V_{\rm BO}+V_{\rm ad}$                  & 44.653 & 45.025 & 45.237(+1.946,-1.706)& 45.243(+1.946,-1.706) \\
$V_{\rm BO}+V_{\rm rel}$                 & 48.793 & 49.244 & 45.234(+1.963,-1.706)& 45.240(+1.946,-1.706) \\
$V_{\rm BO}+V_{\rm ad}+V_{\rm rel}$          & 47.475 & 47.901 & 45.234(+1.946,-1.706)& 45.239(+1.946,-1.706) \\
$V$                              & 47.067 & 47.485 & 45.236(+1.946,-1.706)& 45.243(+1.946,-1.706) \\
$V+V_{\rm ret}$                      & 47.129 & 47.548 & 45.235(+1.946,-1.706)& 45.241(+1.946,-1.706) \\[2mm]
$Hurly$                         & 46.040 & 46.438 & 45.235(+1.946,-1.706)& 45.241(+1.946,-1.706) \\
$HFDHe2$                         & 64.026 & 64.827 & 45.242(+1.946,-1.706)& 45.247(+1.946,-1.706) \\[4mm]
&  &  \qquad\qquad\qquad $^5\Sigma_g^+$  & & \\[2mm]
$V_U$                                      & 49.289 & 50.012 & 48.366(10)& 48.370(9) \\
$V$                                        & 48.542 & 49.234 & 48.365(9) & 48.369(9) \\
$V_L$                                      & 47.827 & 48.489 & 48.363(9) & 48.367(9) \\
$V\!+\!\delta V_{\rm ad}$                     & 47.397 & 48.041 & 48.367(9) & 48.371(9) \\
$V\!+\!\delta V_{\rm rel}$                    & 49.501 & 50.233 & 48.365(9) & 48.369(9) \\
$V\!+\!\delta V_{\rm ad}\!+\!\delta V_{\rm rel}$ & 48.290 & 48.970 & 48.367(9) & 48.371(9) \\[2mm]
$GLD$                                      & 93.189 & 97.265 & 48.410(10)& 48.405(9) \\[2mm]
    \hline\hline
\end{tabular}
\end{center}
{\footnotesize If not stated otherwise the interaction potentials and notation are from Refs.~\cite{PJ,PCJS}.
Calc-1 and Calc-2: derived from  the `zero collision energy' wavefunctions using atomic and nuclear masses, respectively; Calc-3 and Calc-4: evaluated using the `$a$ vs.\ $D_0$' relations with atomic and nuclear masses, respectively. $Hurly$: obtained using the potential of Ref.~\cite{Hurly}. $HFDHe2$: obtained using the potential of Ref.~\cite{Aziz}. $GLD$: obtained using the potential of Ref.~\cite{GLD}.}
\end{table*}

Interestingly, see the middle and bottom panels of Figure 1, Table 1 and Table 2, upon scaling of the interaction potential defined by Eq.~(1), the different potentials all provide closely coinciding relations between the calculated 
properties. Furthermore, as expected from our previous calculations~\cite{SSS}, the same relations are also established for the empirical potentials obtained from fitting to the thermochemical data~\cite{Hurly,Aziz}. This demonstrates the `universality' in two body systems~\cite{Braaten}, where the knowledge of only one of the probed scattering properties, in conjunction with a moderately accurate interaction potential, allows for a quantitative prediction of all the remaining properties.

\begin{figure*}[ht]
\label{fig:fig2}
\includegraphics{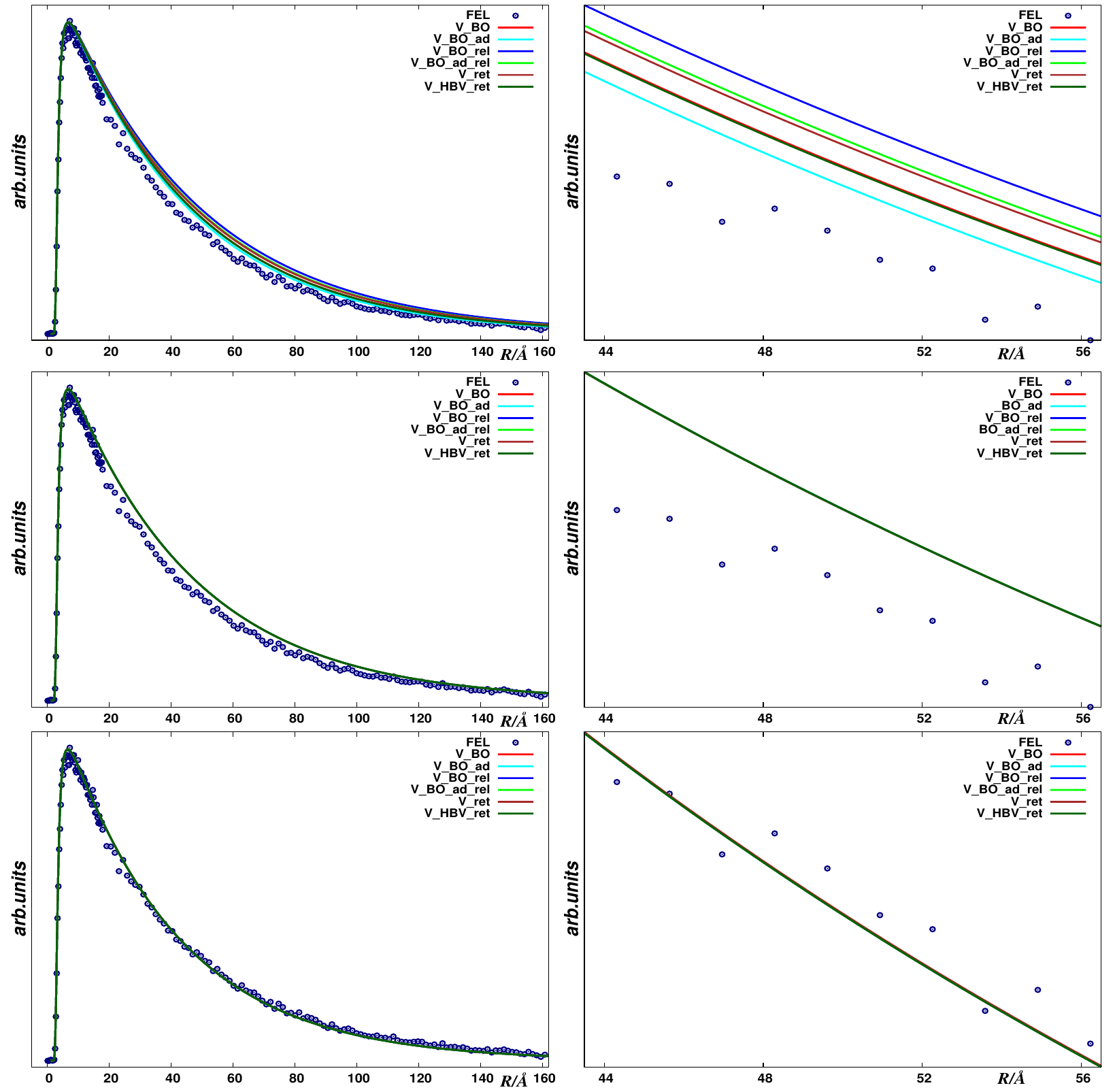}
\caption{
Top panels: Reproduction of the uncorrected FEL data for the $^4$He$_2$ wavefunction ($\Psi$) of Ref.~\cite{Zeller} using the PCJS~\cite{PCJS} and HBV~\cite{HBV} theoretical potentials. Middle panels: The PCJS and HBV potentials scaled to provide the experimental binding energy (36.73 MHz) of Ref.~\cite{Zeller}; $f=1.00040264$, 0.99957216, 1.00237018, 1.00153663, 1.00130907, and 1.00034713 for $V_{\rm BO}$, $V_{\rm BO}\!+\!V_{\rm ad}$, $V_{\rm BO}\!+\!V_{\rm rel}$, $V_{\rm BO}\!+\!V_{\rm ad}\!+\!V_{\rm rel}$, $V\!+\!V_{\rm ret}$, and $V_{\rm HBV}\!+\!V_{\rm ret}$, respectively. Bottom panels: The PJC and HBV potentials scaled to provide the best least squares fit of the data; $f=1.00422756$, 1.00339492, 1.00618642, 1.00534927, 1.00512556, and 1.00417615 for $V_{\rm BO}$, $V_{\rm BO}\!+\!V_{\rm ad}$, $V_{\rm BO}\!+\!V_{\rm rel}$, $V_{\rm BO}\!+\!V_{\rm ad}\!+\!V_{\rm rel}$, $V\!+\!V_{\rm ret}$, and $V_{\rm HBV}\!+\!V_{\rm ret}$, respectively. All calculations were performed using atomic masses. The right hand side panels represent details of the fits.}
\end{figure*}

 The `universality' of the simple scaling defined by Eq.~(1) is also deeply reflected in the fitting of the experimental Coulomb explosion data for the helium dimer wavefunction $\Psi$ of Ref.~\cite{Zeller}. In Figure 2, the top panels illustrate the dispersion of the theoretical $\Psi$ values evaluated using different interaction potentials; the middle panels show the agreement of the theoretical $\Psi$ values obtained with potentials scaled to produce the experimental binding energy $D_0=36.73$ MHz, deduced from experimental data corrected for 
electron-recoil effects; the bottom panels reproduce the uncorrected experimental data using potentials scaled to provide the best least-squares fittings (the effective binding energies corresponding 
to the best reproductions are around 46 MHz, giving insight into the role of electron-recoil effects). 

\begin{figure*}[ht]
\label{fig:fig3}
\includegraphics{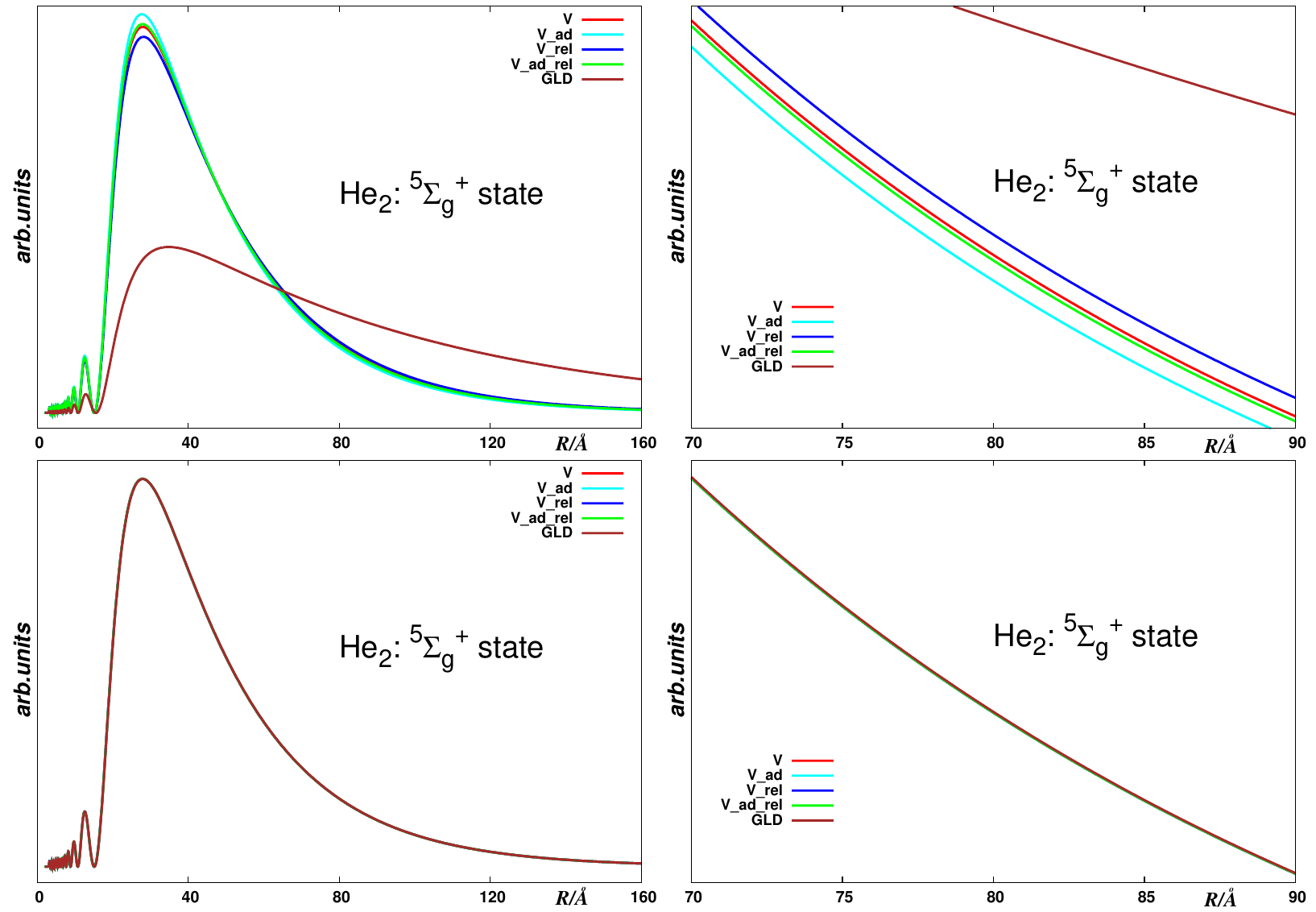}
 \caption{The squares of the $^4$He$_2$ wavefunctions ($|\Psi|^2$) in the $^5\Sigma_g^+$ electronic state evaluated using the original potentials of Refs.~\cite{PJ,GLD} (top panels) and
their versions scaled so that they provide the experimental binding energy of Ref.~\cite{Moal} (bottom panels).}
 \end{figure*}

Generally speaking, the close agreement of the scaled $\Psi$ values is also present in other halo states, for example, the spin-polarized helium atoms in the 2$^3S_1$ metastable state (see Figure 3). Importantly, as seen in the previous figures, the fitted wavefunctions closely coincide over the whole interval of internuclear separations, demonstrating the robustness of the scaling 
approach defined by Eq.~(1) over the whole range of relevant binding energies and scattering lengths. 

\begin{figure*}[ht]
\label{fig:fig4}
\includegraphics{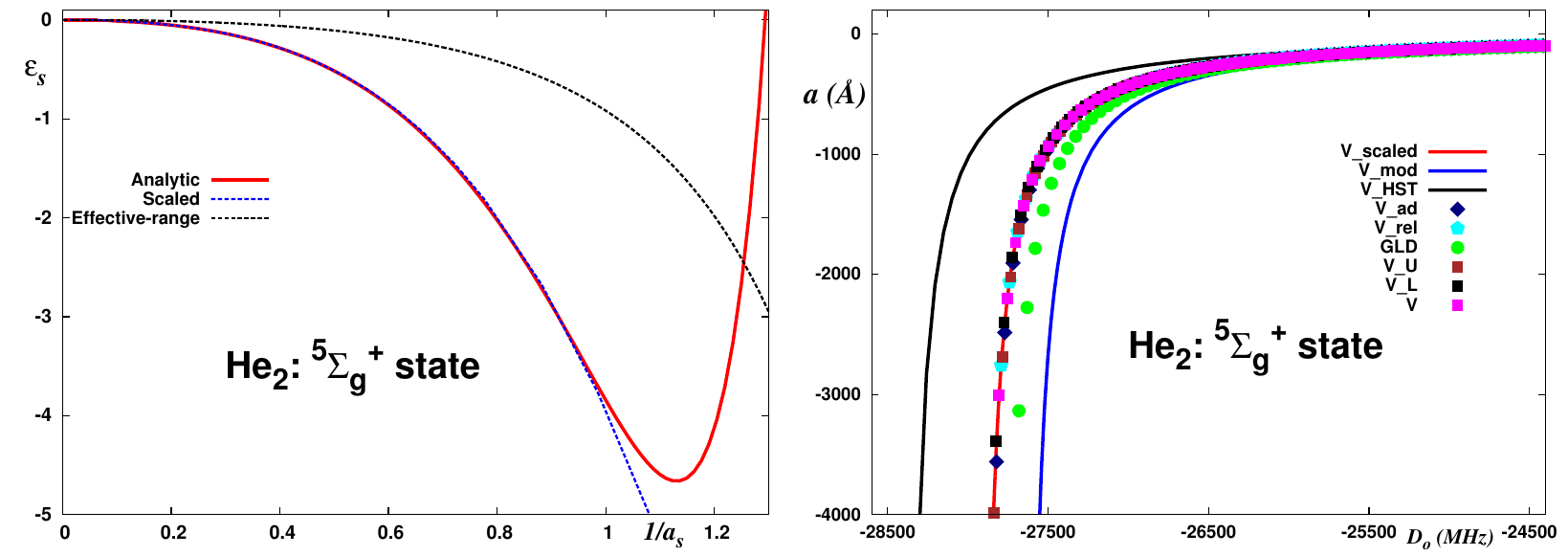}
\caption{Left panel: The scaled binding energy $\epsilon_s=2mDo\beta_{6}^{2}/\hbar^2$ as a function of $1/a_{0s}=\beta_6/a$ ($\beta_6=(2mC_{6}/\hbar^2)^{1/4}$). {\sf Analytic} - results obtained 
using the analytic formula of Ref.~\cite{Gao} (see Eq.~(14) of Ref.~\cite{Gao}). {\sf Scaled} - numerically exact calculations. {\sf Effective range}
- results obtained using $D_0=\hbar^2/2ma^2$. The calculations were performed using  the $V\!+\!\delta V_{\rm ad}\!+\!\delta V_{\rm rel}$ potential energy of Ref.~\cite{PJ}.
Right panel: Comparison of the `$a$ vs.\ $D_0$' relations obtained for the $^5\Sigma_g^+$ state. {\sf V$\_$scaled} - results obtained using the $V\!+\!\delta V_{\rm ad}\!+\!\delta V_{\rm rel}$ theoretical potential of Ref.~\cite{PJ}. {\sf V$\_$mod} - results obtained using the $V\!+\!\delta V_{\rm ad}\!+\!\delta V_{\rm rel}$ potential with discarding the $C_{11}$ and $C_{12}$ contributions. {\sf V$\_$HST} - results obtained
using an effective interaction potential consisting of a hard-sphere with the $-C_6/R^6$ attractive tail (see Eq.(4) of Ref.~\cite{Gao:2004}) and atomic masses.}
 \end{figure*}

In Figure 4, we see that for small binding energies the presented scheme is in excellent agreement with the results obtained using quantum defect theory, which is usually employed in the literature~\cite{Gao,Gao:2004}. However, for higher binding energies associated with scattering lengths tending to $-\infty$, these approaches do not provide reliable asymptotes and thus fail to provide reliable mass sensitivities in these energy regions, particularly for scattering lengths exhibiting a nonlinear and 
discontinuous energy dependence. These regions appear to be particularly promising for probing the mass sensitivity of the scattering properties (see Figure 5), however, as shown in Table 3, the dimensionless scaling of the `global' interaction potential should be adequate for the entire energy region. It should be stressed that only globally accurate potentials appear to be adequate for this critical region.

\begin{figure*}[ht]
\label{fig:fig5}
\includegraphics{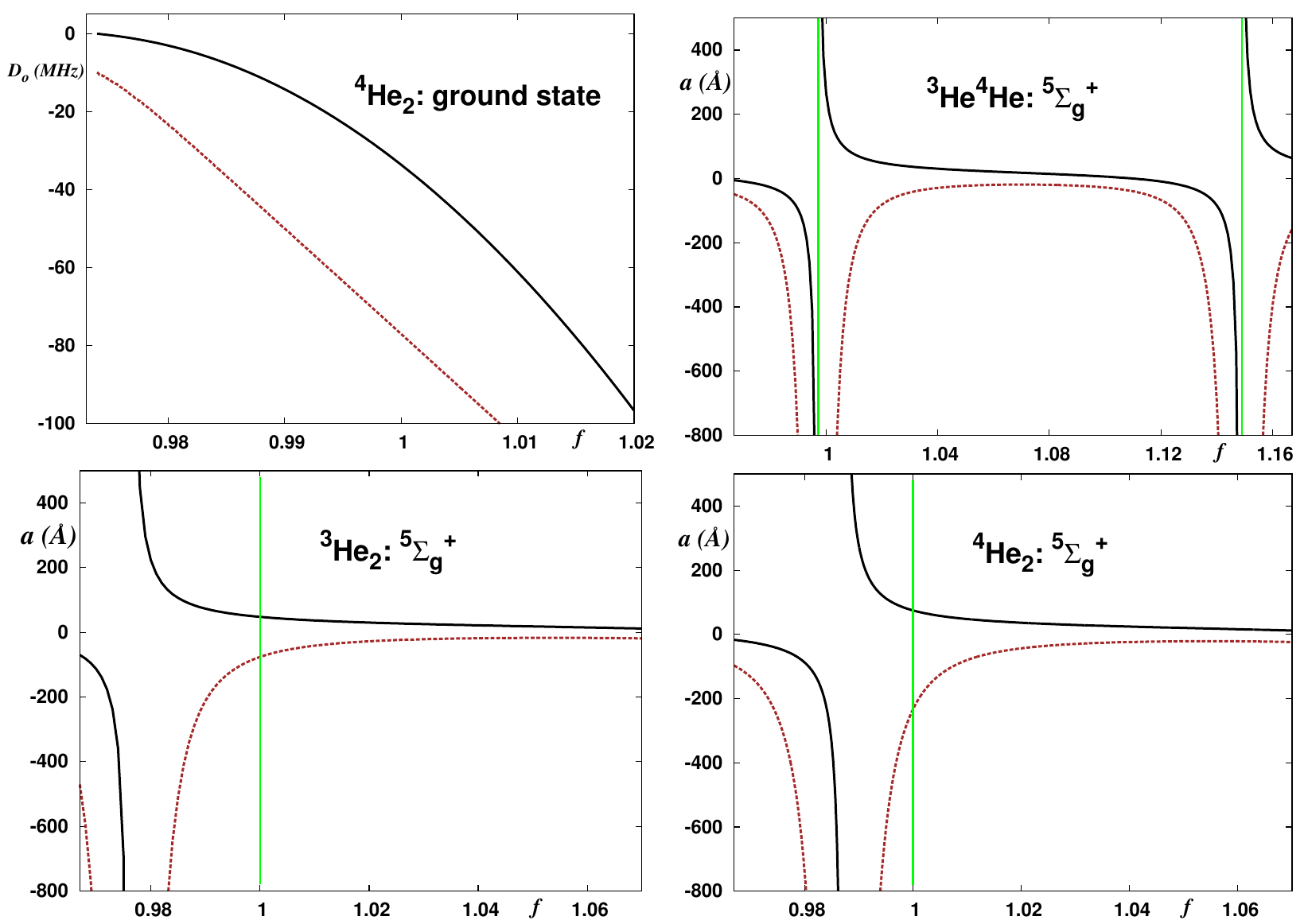}
\caption{Dependence of the binding energy $D_0$ and $s$-wave scattering length $a$ (black lines) of the helium dimers and their derivatives with respect to the molecular reduced mass (dashed brown lines) on the scaling parameter $f$ (derivatives given in arbitrary units). Calculations were performed usingg atomic masses and the $V$ and $V\!+\!\delta V_{\rm ad}\!+\!\delta V_{\rm rel}$ interaction potentials of Ref.~\cite{PCJS} and Ref.~\cite{PJ}, respectively.}
\end{figure*}

\begin{table*}
\caption{\label{tab:scl3} The mass sensitivity coefficients $K_{\beta}$  and $T_{\beta}$ of $^4$He$_2$ in its ground ($^1\Sigma_g^+$) and excited ($^5\Sigma_g^+$) electronic states.}
\vspace*{-3mm}
\begin{center}
\begin{tabular}{c @{\extracolsep{0.30in}}
                c @{\extracolsep{0.22in}}
                c @{\extracolsep{0.22in}}
                c @{\extracolsep{0.22in}}
                c @{\extracolsep{0.12in}}
                c @{\extracolsep{0.12in}}
                c @{\extracolsep{0.22in}}
                c @{\extracolsep{0.22in}}
                c @{\extracolsep{0.22in}}
                c @{\extracolsep{0.22in}}
                }
        \hline\hline\\[-3mm]
Potential  & $K_{1}^a$ & $K_{1}^n$ & $K_{2}^a$ & $K_{2}^n$ & &$T_{1}^a$ & $T_{1}^n$ & $T_{2}^a$ & $T_{2}^n$ \\
        \hline\\[-2mm]
$^1\Sigma_g^+$ &  &  &  & &$^4$He$_2$   & \\[2mm]
$V_{\rm BO}$                         & 66.611 & 67.228 & 65.726 & 65.735& &  -32.35 & -32.66 & -31.91 & -31.91 \\
$V_{\rm BO}+V_{\rm ad}$                  & 64.809 & 65.393 & 65.725 & 65.734& &  -31.45 & -31.75 & -31.92 & -31.92 \\
$V_{\rm BO}+V_{\rm rel}$                 & 71.374 & 72.083 & 65.787 & 65.796& &  -34.73 & -35.09 & -31.94 & -31.95 \\
$V_{\rm BO}+V_{\rm ad}+V_{\rm rel}$          & 69.305 & 69.974 & 65.787 & 65.796& &  -33.70 & -34.03 & -31.94 & -31.95 \\
$V$                              & 68.637 & 69.292 & 65.763 & 65.773& &  -33.37 & -33.69 & -31.93 & -31.94\\
$V+V_{\rm ret}$                      & 68.744 & 69.402 & 65.771 & 65.780& &  -33.42 & -33.75 & -31.93 & -31.94\\[2mm]
$Hurly$                          & 67.035 & 67.660 & 65.772 & 65.781& &  -32.57 & -32.88 & -31.94 & -31.94\\
$HFDHe2$                         & 94.936 & 96.193 & 65.532 & 65.541& &  -46.50 & -47.13 & -31.82 & -31.82\\[2mm]
$^5\Sigma_g^+$ &  &  & & & $^4$He$_2$   & \\[2mm]
$V_U$                                      & 174.11 & 177.74 & 169.49 & 169.51 &  & -64.69 & -66.34 & -62.63 & -62.63 \\
$V$                                        & 170.39 & 173.85 & 169.50 & 169.52 &  & -63.04 & -64.57 & -62.63 & -62.64 \\
$V_L$                                      & 166.83 & 170.14 & 169.51 & 169.53 &  & -61.44 & -62.93 & -62.64 & -62.65 \\
$V\!+\!\delta V_{\rm ad}$                     & 164.64 & 167.86 & 169.49 & 169.51 &  & -60.46 & -61.90 & -62.62 & -62.63 \\
$V\!+\!\delta V_{\rm rel}$                    & 175.19 & 178.86 & 169.50 & 169.52 &  & -65.18 & -66.87 & -62.63 & -62.64 \\
$V\!+\!\delta V_{\rm ad}\!+\!\delta V_{\rm rel}$ & 169.10 & 172.51 & 169.49 & 169.51 &  & -62.45 & -63.99 & -62.62 & -62.63 \\[2mm]
$GLD$                                      & 398.06 & 419.13 & 169.24 & 169.26 &  & -171.3 & -181.7 & -62.25 & -62.25 \\[1mm]
$HST$                                      &  78.55 &  79.22 & 172.28 & 175.83 &  &  -26.39&  -13.51& -70.22 & -72.15 \\[2mm]
 $^5\Sigma_g^+$ &  &  & & & $^3$He$_2$   & \\[2mm]
 $V$                                        & 94.31 & 95.64 & 94.05 & 95.37 & & -31.88 & -32.42 & -31.78& -32.30  \\
 $V\!+\!\delta V_{\rm ad}$                     & 92.60 & 93.88 & 94.04 & 95.37 & & -31.21 & -31.72 & -31.78& -32.30 \\
 $V\!+\!\delta V_{\rm rel}$                    & 95.69 & 97.06 & 94.05 & 95.37 & & -32.43 & -32.98 & -31.78& -32.30 \\
 $V\!+\!\delta V_{\rm ad}\!+\!\delta V_{\rm rel}$ & 93.93 & 95.25 & 94.04 & 95.37 & & -31.73 & -32.26 & -31.78& -32.30 \\[2mm]
$GLD$                                       & 134.3 & 137.2 & 94.10 & 95.42 & & -48.79 & -50.04 & -31.80& -32.32 \\[1mm]
$HST$                                       & 65.92 & 66.54 & 90.82 & 92.05 & & -22.10 & -22.31 & -31.78& -32.31 \\[2mm]
 $^5\Sigma_g^+$ &  &  & & & $^3$He$^4$He   & \\[2mm]
 $V$                                        & 668.4 & 746.1 & 653.4 & 727.3 & & -306.3 & -344.6 & -298.8& -335.4 \\
 $V\!+\!\delta V_{\rm ad}$                     & 579.9 & 636.8 & 653.2 & 727.0 & & -262.4 & -290.6 & -298.7& -335.2 \\
 $V\!+\!\delta V_{\rm rel}$                    & 759.6 & 863.6 & 653.4 & 727.3 & & -351.6 & -402.4 & -298.8& -335.3 \\
 $V\!+\!\delta V_{\rm ad}\!+\!\delta V_{\rm rel}$ & 646.8 & 719.1 & 653.2 & 727.0 & & -295.6 & -331.3 & -298.7& -335.2 \\[2mm]
 $GLD$                                      & 17.51 & 17.55 & 652.6 & 726.1 & &  248.3 &  233.4 & -305.8& -345.3 \\[1mm]
 $HST$                                      &127.3  & 129.5 & 492.8 & 532.9 & & -47.93 & -49.02 & -298.8& -335.5 \\[2mm]
    \hline\hline
\end{tabular}
\end{center}
{\footnotesize If not stated otherwise the interaction potentials and notation are from Refs.~\cite{PJ,PCJS}. $D_0=36.73$ MHz~\cite{Zeller} and 91.35 MHz~\cite{Moal} for the ground and excited state of $^4$He$_2$, respectively. $K_{1}^a$/$T_{1}^a$ and $K_{1}^n$/$T_{1}^n$: derived from the `zero collision energy' wavefunctions using atomic and nuclear masses, respectively; $K_{2}^a$/$T_{2}^a$ and $K_{2}^n$/$T_{2}^n$: evaluated using `$a$ vs.\ $D_0$' with atomic and nuclear masses, respectively. $Hurly$: obtained using
the potential of Ref.~\cite{Hurly}. $HFDHe2$: obtained using the potential of Ref.~\cite{Aziz}. $GLD$: obtained using the potential of Ref.~\cite{GLD}. $HST$: Calculations performed using the $V_{\rm HST}$ effective potential of Ref.~\cite{Gao:2004} with $r_0=3.8$~{\AA} and $C_6=3276.68$ {\it a.u.}~\cite{PJ}. In the case of $^3$He$_2$ and $^3$He$^4$He, where $D_0$ is not available, $V_{\rm HST}$ was scaled so that it provides the same $T_2^a$ as its {\it ab initio} counterpart $V$.}
\end{table*}

\section{Conclusions}

 The scattering properties of diatomic halo states can be extracted from one-parameter only $f$-relationships. These relations were derived from numerically exact solutions of the radial Schr\"odinger equation for a set of effective potentials, obtained by a multiplicative scaling of the `generic' interaction potential. The calculated scattering properties are as accurate as their unscaled $f=1$ counterparts, which are derived from the best available interaction potentials. Furthermore, the predicted values appear somewhat independent of the underlying potential, suggesting that any moderately accurate potential can be utilized in the presented approach. As anticipated, the mass sensitivities of the scattering lengths and of the related properties grow significantly when increasing the classically forbidden part of the halo state wavefunction. For example, the sensitivities of the delocalized state of $^4$He$^3$He are about one order of magnitude larger than those relating to a less delocalized state of $^3$He$_2$. 

 Most importantly, a single-parameter scaling based on Eq.~(1) allows for the close fitting of the square of the experimental halo state wavefunction $|\Psi|^2$, thus reflecting the universality of quantum diatomic halo states of diatomic systems (see, e.g., Ref.~\cite{Braaten}). This universality means that all the parameters
characterizing the low-energy scattering of atoms can be determined with high precision if only one of these characteristics is accurately known. Moreover, scattering properties and their mass sensitivities, which appear promising for investigating the stability of the electron-to-proton mass ratio $\beta$, can be fully determined. All one requires is a potential energy curve of moderate accuracy and an accurate experimental value for a pertinent scattering property. In principle, unlike the semiclassical~\cite{Boisseau} or quantum defect theories~\cite{Gao:r4,Idziaszek} which can be viewed as the standard alternatives for deriving the `$a$ vs.\ $D_0$' relations, the scaling described by Eq.~(1) can be used straightforwardly for any type of long-range potential asymptote. 

 Currently, the best (spectroscopic) measurements of the scattering length can probe a variation of $\beta$ at the level of 10$^{-13}$--10$^{-16}$ yr$^{-1}$~\cite{Chin}. Given that the scattering phase shift $\delta_0$ can be measured with a precision that yields scattering lengths with 1 ppm accuracy~\cite{Hart}
(see also Ref.~\cite{Gacesa}), one can expect to investigate temporal variations of $\beta$ at the level of 10$^{-15}$--10$^{-18}$ yr$^{-1}$. The presented approach can cope with this accuracy by providing accurate mass sensitivities even in cases where a lack of spectral data prevents the use of standard approaches. However, it should be noted that the accuracy of the experimental setup will dictate future investigations of $\beta$, and large sensitivity coefficients may not always be transferable to experiment.

\begin{acknowledgments}
We thank Stefan Zeller and his coauthors for the Coulomb explosion data. This work was a part of the research project RVO:61388963 (IOCB) and was supported by the Czech Science Foundation (grant P106/12/G015). A.O.\ gratefully acknowledges a fellowship from the Alexander von Humboldt Foundation and support from the \emph{Deutsche Forschungsgemeinschaft} (DFG) through the excellence cluster ``The Hamburg Center for Ultrafast Imaging -- Structure, Dynamics and Control of Matter at the Atomic Scale'' (CUI, EXC1074).
\end{acknowledgments}

\clearpage
\newpage

\bibliographystyle{apsrev}

\end{document}